\documentstyle[prb,aps,twocolumn,floats,amssymb,amsfonts,epsfig]{revtex}

\begin{document}
\draft \flushbottom

\wideabs{

\title{
Gorkov equations for a pseudo-gapped high temperature superconductor
}

\author{
B. Giovannini\cite{email} and C. Berthod
}

\address{
DPMC, Universit\'e de Gen\`eve, 24 Quai Ernest-Ansermet,
1221 Gen\`eve 4, Switzerland
}

\date{August 3, 2000}
\maketitle

\begin{abstract}

A theory of superconductivity based on the two-body Cooperon propagator is
presented. This theory takes the form of a modified Gorkov equation for the
Green's function and allows one to model the effect of local superconducting
correlations and long range phase fluctuations on the spectral properties of
high temperature superconductors, both above and below $T_c$. A model is
proposed for the Cooperon propagator, which provides a simple physical picture
of the pseudo-gap phenomenon, as well as new insights into the doping
dependence of the spectral properties. Numerical calculations of the density of
states and spectral functions based on this model are also presented, and
compared with the experimental STM and ARPES data. It is found, in particular,
that the sharpness of the peaks in the density of states is related to the
strength and the range of the superconducting correlations and that the
apparent pseudo-gap in STM and ARPES can be different, although the underlying
model is the same.

\end{abstract}


}

\section{Introduction}

The anomalous properties of high temperature superconductors (HTS) have been
the object of many investigations over the last ten years.\cite{Timusk_99}
Among these anomalous properties, the pseudo-gap phenomenon (a substantial
decrease of the one particle density of states near the Fermi energy in the
normal state below a certain temperature $T^*$) has been studied by several
experimental techniques. In tunneling spectroscopy experiments, in particular,
the pseudo-gap manifests itself as a gap in the excitation
spectrum.\cite{Renner_98} On the theoretical front, many competing models have
been proposed.\cite{Roddick_95,Emery_97,Anderson_97,Randeria_89,%
Schrieffer_95,Chubukov_96,Lee_97,Ranninger_95,Geshkenbein_97,Perali_99} One of
the popular interpretations of the pseudo-gap is that superconductivity forms
{\em locally\/}\ at $T^*$, but the phases of distant superconducting
``droplets'' remain incoherent until the superconducting transition temperature
$T_c$ is reached.\cite{Roddick_95,Emery_97} This view is supported by an
increasing evidence that the pseudo-gap phenomenon is intimately connected to
the underlying superconducting phase, mainly because the $d$-wave symmetry of
the pseudo-gap is the same as that of the gap in the superconducting
phase.\cite{Loeser_96,Ding_96} The phase fluctuation model of the pseudo-gap
state is different from the usual theory of superconducting fluctuations,
because the latter is only valid near $T_c$, and involves both size and phase
fluctuations.\cite{Curty_00}

A theory of the pseudo-gap above $T_c$ has also consequences below $T_c$. In
particular, it is connected to the character of the excitations responsible for
destroying superconductivity. Different mechanisms (thermal phase fluctuations,
quantum phase fluctuations, nodal quasiparticles) may all contribute to the
properties in the superconducting state, and these contributions may be of
varying importance if one considers low temperatures or temperatures near
$T_c$.\cite{Paramekanti_00} One may add that the clue to a theory of high
temperature superconductivity will go through the explanation of detailed
properties, like the absence of quasiparticles above $T_c$ and their appearance
below $T_c$,\cite{Kaminski_00} or the anomalous properties of the density of
states in vortices.\cite{Renner_98b}

Within the BCS theory, the inclusion of phase fluctuations in a droplet model
must be done in two steps. First, one introduces a local BCS gap, with a phase,
and then one must somehow analyze the phase fluctuations in a second analytical
step distinct from the first. The theory of Franz and Millis\cite{Franz_98}
gives an example of this type of approach. Starting from the form of the
Green's function in a uniform superflow, they extend it semi-classically to
non-uniform situations assuming slow spatial variations of the superfluid
velocity. This Green's function is then averaged over a Gaussian distribution
of velocity fluctuations, which relates the result to a correlation function of
the velocities. A similar approach has been used by Kwon and
Dorsey,\cite{Kwon_99} who treat the coupling to the fluctuating phase using a
self-consistent perturbation theory. As emphasized by Geshkenbein {\it et
al.}\cite{Geshkenbein_97} and Randeria,\cite{Randeria_97} strong pairing
correlations should however be incorporated at the basis of any model of the
pseudo-gap regime.

A systematic theory of the effect of phase fluctuations on the density of
states (and other properties) in HTS, above and below $T_c$, must start by
putting the phase-phase correlation function at the core of the theory for
superconductivity, at the same level as the size of the local gap. This program
implies that one avoids developing the theory of superconductivity by defining
a gap function and an anomalous propagator. This is equivalent to writing the
BCS theory in a particle number conserving scheme ({\it i.e.\/} in the
canonical ensemble, avoiding the definition of an anomalous amplitude between
states with different particle numbers). This theory has actually been written
down forty years ago by Kadanoff and Martin\cite{Kadanoff_61} (KM), and
rediscovered by others, in particular for the discussion of Josephson
arrays.\cite{Weiss_82} The KM theory is based on the two-body Cooperon
propagator, and describes quite naturally the effect of phase fluctuations.
This theory has already been applied to the HTS in a nice series of papers by
Levin {\it et al.},\cite{Janko_97} but with a different interpretation (not
related to phase fluctuations), a different focus, and a different formalism
than in our work. In this paper we rewrite the basic KM equations for the case
of a lattice Hamiltonian, we show again how the standard BCS theory is
recovered with a straightforward approximation for the two-body Cooperon
correlation function, and we then derive the basic equations for a pseudo-gap
state, in particular the equivalent Gorkov equations which have to be used in
the calculation of vortex states.

\section{Kadanoff-Martin equations in a lattice model}
\label{sect:KM}

We consider the lattice Hamiltonian
	\begin{equation}\label{eq:hamiltonian}
		{\cal H}=\sum_{ij\sigma} t^{ }_{ij}\,
		c^{\dagger}_{i\sigma}c^{ }_{j\sigma}+
		\sum_{ij} V^{ }_{ij}\,b^{\dagger}_i b^{ }_j.
	\end{equation}
In this, $c^{\dagger}_{i\sigma}$ creates an electron at site $i$ and
$b^{\dagger}_i=c^{\dagger}_{i\downarrow}c^{\dagger}_{i\uparrow}$ creates a
Cooper pair at site $i$. We assume that $t$ and $V$ are symmetric and real. The
usual Gorkov equations can then be written as a single equation:
	\begin{equation}\label{eq:gorkov1}
		\big[{\cal G}^0(\omega_n)\big]^{-1}{\cal G}(\omega_n) =
		\openone+\tilde{\Sigma}(\omega_n)\,{\cal G}(\omega_n),
	\end{equation}
where ${\cal G}_{ij}(\tau) = -\langle T_{\tau}\{c^{ }_{i\uparrow}(\tau)
c^{\dagger}_{j\uparrow}(0)\}\rangle$ and
	\begin{equation}\label{eq:sigmatilde}
		\tilde{\Sigma}_{ij}(\omega_n) = -\sum_{r_1r_2}
		V^{ }_{ir_1}B^{ }_{r_1}B^{\star}_{r_2}
		V^{ }_{r_2j}{\cal G}_{ji}^0(-\omega_n).
	\end{equation}
$\big[{\cal G}^0(\omega_n)\big]^{-1}_{ij} = i\omega_n\delta_{ij}-t_{ij}$ is the
free Green's function, $\omega_n$ are the odd Matsubara frequencies, and
$B_{\ell}=\left\langle b_{\ell}\right\rangle$. The notation $\big[{\cal
G}^0\big]^{-1}{\cal G}$ implies matrix multiplication in the $\{i,\,j\}$ space.
These equations are supplemented by the self-consistent equation for $B_i$ (or
$\Delta_i$) and the self-consistent equation relating the number of particles
$N$ to the chemical potential. The equation for $B_i$ is
	\begin{equation}\label{eq:scB1}
		B_i=-{\cal F}_{ii}(0^+)=-\frac{1}{\beta}\sum_{\omega_n}
		{\cal F}_{ii}(\omega_n)\,e^{-i\omega_n0^+}
	\end{equation}
with ${\cal F}^{\star}_{ij}(\tau) = -\langle T_{\tau}\{
c^{\dagger}_{i\downarrow}(\tau) c^{\dagger}_{j\uparrow}(0)\}\rangle$.

The Kadanoff-Martin correlation function description of superconductivity
consists simply (after a long and thorough discussion of higher order
correlation functions) in replacing $\tilde{\Sigma}$ in Eq.~(\ref{eq:gorkov1})
by
	\begin{equation}\label{eq:sigmaKM}
		\Sigma_{ij}(\omega_n)=-\frac{1}{\beta}\sum_{\omega_m}\sum_{r_1r_2}
		V^{ }_{ir_1}L^{ }_{r_2r_1}(\omega_n+\omega_m)V^{ }_{r_2j}
		{\cal G}_{ji}^0(\omega_m)
	\end{equation}
where $L_{r_2r_1}(\tau)=\left\langle T_{\tau}\left\{b^{ }_{r_1}(\tau)
b^{\dagger}_{r_2}(0)\right\}\right\rangle$ is the Cooperon propagator. The
self-consistent equation for $B$ is replaced by the equations
	\begin{mathletters}
	\label{eq:scL}
	\begin{eqnarray}\label{eq:scLaTc}
		\nonumber && L_{r_2r_1}(\Omega_m)=-\frac{1}{\beta}\sum_{\omega_n}
		{\cal G}^{ }_{r_2r_1}(\Omega_m+\omega_n){\cal G}^0_{r_1r_2}(-\omega_n)\\
		&&\quad -\frac{1}{\beta}\sum_{\omega_n}\sum_{ij}
		{\cal G}^{ }_{ir_2}(\Omega_m+\omega_n)
		{\cal G}^0_{ir_2}(-\omega_n)V^{ }_{ji}L^{ }_{jr_1}(\Omega_m)
	\end{eqnarray}
for $T>T_c$, and
	\begin{eqnarray}\label{eq:scLbTc}
		\nonumber && L_{r_2r_1}(\Omega_m) = \\
		&&\quad -\frac{1}{\beta}\sum_{\omega_n}\sum_{ij}
		{\cal G}^{ }_{ir_2}(\Omega_m+\omega_n)
		{\cal G}^0_{ir_2}(-\omega_n)V^{ }_{ji}L^{ }_{jr_1}(\Omega_m)
	\end{eqnarray}
	\end{mathletters}
for $T<T_c$, where $\Omega_m$ are the even Matsubara frequencies. The fact that
the inhomogeneous term in the ladder equation for $L$ has to be dropped when
calculating the order parameter in the condensed state is not much commented
upon in the original paper of KM, but it is related to the range of
$L_{r_2r_1}(\Omega_m)$, which is finite above $T_c$ and infinite below $T_c$
(see below). In fact, the quantity $\sum_{r_1r_2}L_{r_2r_1}(\Omega_m)$ is of
the order $N^2$ below $T_c$ (where $N$ is the number of sites) whereas the
corresponding sum of the inhomogeneous term in Eq.~(\ref{eq:scLaTc}) is only of
order $N$. The usual BCS theory is recovered by setting
	\[
		L_{r_2r_1}(\tau) = B^{ }_{r_1}B^{\star}_{r_2}, \quad
		L_{r_2r_1}(\Omega_m) =
		\beta B^{ }_{r_1}B^{\star}_{r_2}\delta_{\Omega_m,0}
	\]
in Eq.~(\ref{eq:sigmaKM}). The self-consistent equation Eq.~(\ref{eq:scLbTc})
for $L$ then goes over into the self-consistent equation Eq.~(\ref{eq:scB1})
for $B$, and clearly Eq.~(\ref{eq:sigmaKM}) goes into
Eq.~(\ref{eq:sigmatilde}). In the BCS framework, the Thouless criterion for
$T_c$ becomes the gap equation for $T<T_c$.

The KM description of superconductivity, which is entirely based on the
properties of the function $L$, is thus seen to be a natural starting point if
one wants to introduce explicitly local order and phase fluctuations in the
physical description of high temperature superconductors. This is done in the
next section.

\section{Phenomenological description of a pseudo-gapped superconductor}

Our fundamental assumption is that Eq.~(\ref{eq:sigmaKM}) is generally valid,
in the sense that it expresses in general the single particle Green's function
in terms of the Cooperon propagator, regardless of the model or the
approximations involved in calculating this propagator. Our purpose in this
paper is to explore the experimental consequences of a simple heuristic form
for $L$, which is the translation of the physical picture presented in the
Introduction. For $T>T_c$, we write:
	\begin{mathletters}
	\label{eq:L}
	\begin{equation}\label{eq:LTgtTc}
		L_{r_2r_1}(\tau)=|B^0|^2\,R(r_{12}/\varrho_0)+
		|B^1|^2\,F(r_2-r_1,\,\tau)
	\end{equation}
with $r_{12}=|r_2-r_1|$ and $R(x)$ some cutoff function which vanishes rapidly
for $x>1$. This equation expresses the fact that there are strong
superconducting correlations at the scale $\varrho_0$, represented by a finite
value of $B^0$, going to zero gradually at a temperature $T^*$. The strength of
the superconducting correlations between ``droplets'' is represented by an
amplitude $B^1$ and some function $F$ describing essentially the correlations
in an $XY$ model above the Kosterlitz-Thouless (KT) transition. Both $B^1$ and
$F$ will be temperature dependent. As $T$ approaches $T_c$ from above (we
identify $T_c$ with the KT transition temperature), the correlation length of
$F$ diverges and $F$ approaches 1. For $T<T_c$, we write:
	\begin{equation}\label{eq:LTltTc}
		L_{r_2r_1}(\tau)=|B^0|^2\,R(r_{12}/\varrho_0)+
		B^1_{r_1}{B^1_{r_2}}^{\star},
	\end{equation}
	\end{mathletters}
where only the amplitude $B^1$ carries a phase. The assumption here is that
short range correlations have a strong incoherent part, even in the
superconducting state. When introduced into the equation for the self-energy,
Eq.~(\ref{eq:LTltTc}) for $L$ means that the self-energy in the superconducting
state is the sum of a coherent and an incoherent part; this appears to be the
case in some recent calculations based on a fermion-boson
model.\cite{Barnes_99} Inspection of Eqs.~(\ref{eq:LTltTc}) and (\ref{eq:scL})
shows that if, in the ladder approximation for a homogeneous system, it turns
out that $L_{r_2r_1}$ is the sum of a constant term and a term of finite range,
then the constant term will obey Eq.~(\ref{eq:scLbTc}), whereas the finite
range term will obey Eq.~(\ref{eq:scLaTc}).

Eqs.~(\ref{eq:L}) translate into the following equation for ${\cal G}$:
	\begin{equation}\label{eq:new_gorkov1}
		\big[\tilde{\cal G}^0(\omega_n)\big]^{-1}{\cal G}(\omega_n) =
		\openone+\Sigma^1(\omega_n)\,{\cal G}(\omega_n),
	\end{equation}
where
	\begin{eqnarray}
		\nonumber && \big[\tilde{\cal G}^0(\omega_n)\big]^{-1}_{ij}=
		\big[{\cal G}^0(\omega_n)\big]^{-1}_{ij} \\
		&&\qquad +\sum_{r_1r_2}V^{ }_{ir_1}|B^0|^2\,R(r_{12}/\varrho_0)\,
		V^{ }_{r_2j}{\cal G}^0_{ji}(-\omega_n)
	\end{eqnarray}
and
	\begin{mathletters}
	\begin{eqnarray}
		\label{eq:sigma1_above}\nonumber
		\Sigma^1_{ij}(\omega_n)&=&-\frac{1}{\beta}\sum_{\omega_m}
		\sum_{r_1r_2}V^{ }_{ir_1}|B^1|^2 F(r_2-r_1,\,\omega_n+\omega_m)
		\\[-0.8em] &&\hspace*{7.2em} \times
		V^{ }_{r_2j}{\cal G}^0_{ji}(\omega_m),\; T\!>\!T_c \\[0.5em]
		\label{eq:sigma1_below}
		\Sigma^1_{ij}(\omega_n)&=&{\displaystyle -\!\!\sum_{r_1r_2}
		V^{ }_{ir_1}B^1_{r_1}{B^1_{r_2}}^{\!\star}
		V^{ }_{r_2j}{\cal G}^0_{ji}(-\omega_n)},\; T\!<\!T_c.
	\end{eqnarray}
	\end{mathletters}
Inspection of Eq.~(\ref{eq:sigma1_below}), together with
Eq.~(\ref{eq:new_gorkov1}), shows that below $T_c$, a pseudo-gapped
superconductor obeys the following modified Gorkov equations:
	\begin{mathletters}
	\begin{eqnarray}
		\left\{\big[\tilde{\cal G}^0(\omega_n)\big]^{-1}
		{\cal G}(\omega_n)\right\}_{\!ij}
		+\!\sum_{\ell} V^{ }_{i\ell}B_{\ell}^1
		\tilde{\cal F}_{ij}^{\star}(\omega_n) &=& \delta_{ij} \\
		\left\{\big[{\cal G}^0(-\omega_n)\big]^{-1}\!
		\tilde{\cal F}^{\star}(\omega_n)\right\}_{\!ij}\!\!
		-\!\sum_{\ell}V^{ }_{\ell i}{B_{\ell}^1}^{\star}
		{\cal G}^{ }_{ij}(\omega_n) &=& 0.
	\end{eqnarray}
	\end{mathletters}

Quantum Monte Carlo (QMC) calculations of the pairing correlations were
recently reported for the attractive Hubbard model at zero
temperature.\cite{Guerrero_00} Although these results are restricted to short
distances ($\sim$ 6--10 lattice sites) we tentatively connect our model to the
QMC calculations with the following arguments. For the system sizes considered
in the QMC calculations (typically $14\times14$ sites) the correlation function
for the largest distance in the system has not converged to its asymptotic
value. We attribute the slow decrease of the correlations at intermediate
distances (see inset of Fig.~4 in Ref.~\onlinecite{Guerrero_00}) to a large
value of $\varrho_0$ with respect to the system size. The results of
Ref.~\onlinecite{Guerrero_00} also show that the strength of the pairing
correlations at intermediate distances increases, and differs increasingly from
the BCS result, as the Hubbard interaction $U/t$ increases. Closer inspection
of the data in Fig.~3 of Ref.~\onlinecite{Guerrero_00} indicates that the ratio
of the BCS to the QMC correlations at intermediate distances is also an
increasing function of $U/t$. The simplest BCS approximation to
Eq.~(\ref{eq:LTltTc}) is to replace the cutoff function $R$ by $1$, describing
correlations which are independent of $r_1$ and $r_2$ (in a homogeneous system,
the second term of Eq.~(\ref{eq:LTltTc}) is a constant $|B^1|^2$). With this
approximation, one can account for the above trends by assuming that both $B^0$
and the ratio $B^0/B^1$ increase as $U/t$ increases. Finally, we shall include
in our numerical calculations the ``onsite'' correlations found in
Ref.~\onlinecite{Guerrero_00} for distances within two lattice spacings, by
adding a term
	\begin{equation}\label{eq:Los}
		L^{\text{os}}_{r_2r_1}=\left(|B^0|^2+|B^1|^2\right)e^{-2\,r_{12}/a}
	\end{equation}
to the model Eqs.~(\ref{eq:L}), where $a$ is the lattice parameter. We find,
however, that this correction has a negligible impact on the spectral
functions, and could equally be dropped without changing the results presented
below.

\section{Numerical results}

We now use the general equations derived in Section~\ref{sect:KM}, together
with the model Eqs.~(\ref{eq:L}) and (\ref{eq:Los}), to
calculate the temperature dependence of the density of states and spectral
functions in a homogeneous system. The calculations are compared with the STM
and ARPES experimental results for BSCCO. In order to reduce the number of
adjustable parameters, we take for the correlation function $F(\bbox{r},\,\tau)
= \exp[-r/\xi(T)]$, which describes time-independent phase-phase correlations
above $T_c$. The correlation length $\xi$ is equal to
$\varrho_1\alpha\exp[b/\sqrt{(T-T_c)/J}]$, with $\alpha\approx 0.21$, $b\approx
1.73$, and $J\approx T_c/0.89$.\cite{Gupta_92} The length-scale $\varrho_1$ is
the lattice parameter of the effective $2D-XY$ model describing phase
fluctuations. We found that the main features in the spectral properties above
$T_c$ are rather insensitive to the details of the correlation function. We
have explicitly checked this point by comparing different functions $F$, in
particular functions which give a better description of the correlations in the
$XY$ model. The cutoff function $R$ is modeled as $\exp(-r/\varrho_0)$.

\begin{figure}[bt!]
\leavevmode\begin{center}\epsfxsize8cm\epsfbox{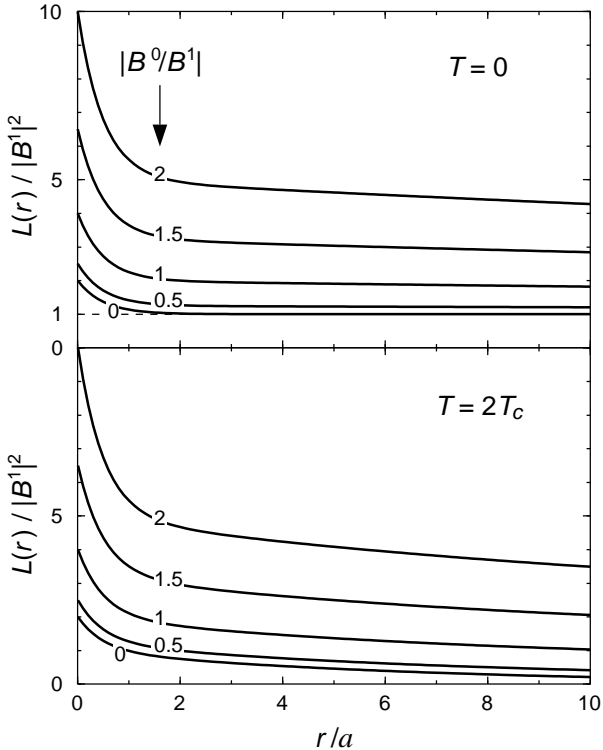}\end{center}
\caption{
Model two-body Cooperon correlation function at temperatures below and above
$T_c$. The increase of the correlations for $r\lesssim a$ is due to
$L^{\text{os}}$ given in Eq.~(\ref{eq:Los}). Below $T_c$, $L(r)$ converges to
the finite asymptotic value $|B^1|^2$ at distances of the order $\varrho_0$ if
$B^0>0$ and of the order $a$ if $B^0=0$. Above $T_c$, the range of $L(r)$ is
finite. If $B^0>0$, this range is given by $\max(\varrho_0,\,\xi(T))$ while if
$B^0$ vanishes it is given by $\xi(T)$.
}\label{fig:L}
\end{figure}

The four parameters $\varrho_0$, $\varrho_1$, $B^0$, and $B^1$ are chosen to
achieve good agreement with the experimental results. We use the value
$\varrho_0=50a$; if the first term in the right-hand side of
Eq.~(\ref{eq:LTltTc}) is the dominant one ($B^0>B^1$), we found that a
relatively large value of $\varrho_0$ is needed in order to obtain well
developed coherence peaks in the zero temperature density of states. In
addition, we see that, according to the previous discussion, $\varrho_0$ must
be large with respect to $\sim14a$. The parameter $\varrho_1$ controls the
temperature evolution of the spectral functions above $T_c$, and takes the
value $\varrho_1=5a$. The larger $\varrho_1$, the wider the temperature region
above $T_c$ in which finite range phase coherence contributes to the
pseudo-gap. The amplitude $B^0$ is adjusted to fix the gap energy to
$\sim40$~meV. Finally, the ratio $B^0/B^1$ is varied in order to control the
relative importance of short range superconducting correlations and long-range
phase fluctuations. The behavior of the resulting Cooperon propagator
$L(\bbox{r})$ is illustrated in Fig.~\ref{fig:L} for temperatures below and
above $T_c$ and for different values of $|B^0/B^1|$.

For a translationally invariant system and our model Cooperon propagator,
Eq.~(\ref{eq:sigmaKM}) can be recast as
	\begin{equation}\label{eq:sigma_k}
		\Sigma(\bbox{k},\,\omega_n)=\frac{1}{(2\pi)^2}\int_{\text{BZ}}
		\frac{V^2(\bbox{q})L(\bbox{q})d\bbox{q}}
		{i\omega_n+\varepsilon_{\bbox{q}-\bbox{k}}}
	\end{equation}
where $V(\bbox{q})=V_0+2V_1(\cos q_xa+\cos q_ya)$, $L(\bbox{q})$ is the Fourier
transform of $L(\bbox{r})$, and $\varepsilon_{\bbox{k}}$ is the free
dispersion. Here, $V_0$ and $V_1$ are the onsite and nearest-neighbor
potentials, respectively, and we neglect next-nearest neighbor interactions; we
assume $V_1=V_0/4$ in all of our calculations. For the dispersion, we use a
tight-binding expression which fits the BSCCO Fermi surface and corresponds to
a bandwidth of 2~eV.\cite{Schmalian_97} The self-energy at real frequencies is
evaluated by making the analytic continuation
$i\omega_n\rightarrow\omega^+=\omega+i0^+$ in Eq.~(\ref{eq:sigma_k}) and
discretizing the BZ integral.\cite{FFT} The spectral function is then
calculated according to $A(\bbox{k},\,\omega)=
-\frac{1}{\pi}\text{Im}\big\{\!\left[\omega^+
-\varepsilon_{\bbox{k}}-\Sigma(\bbox{k},\,\omega^+)\right]^{-1}\!\big\}$, and
the density of states is $N(\omega)\propto
\int_{\text{BZ}}A(\bbox{k},\,\omega)\,d\bbox{k}$. It is easy to check from
Eq.~(\ref{eq:sigma_k}) that, if $L(\bbox{q})>0$ --- a condition obeyed by our
model --- then the Green's function is analytic in the upper half of the
complex plane, the spectral function $A(\bbox{k},\,\omega)$ is positive, and
the Green's function goes to zero as $\omega^{-1}$ for
$|\omega|\rightarrow\infty$.

\subsection{Scanning tunneling spectroscopy}

\begin{figure}[tb!]
\leavevmode\begin{center}\epsfxsize7cm\epsfbox{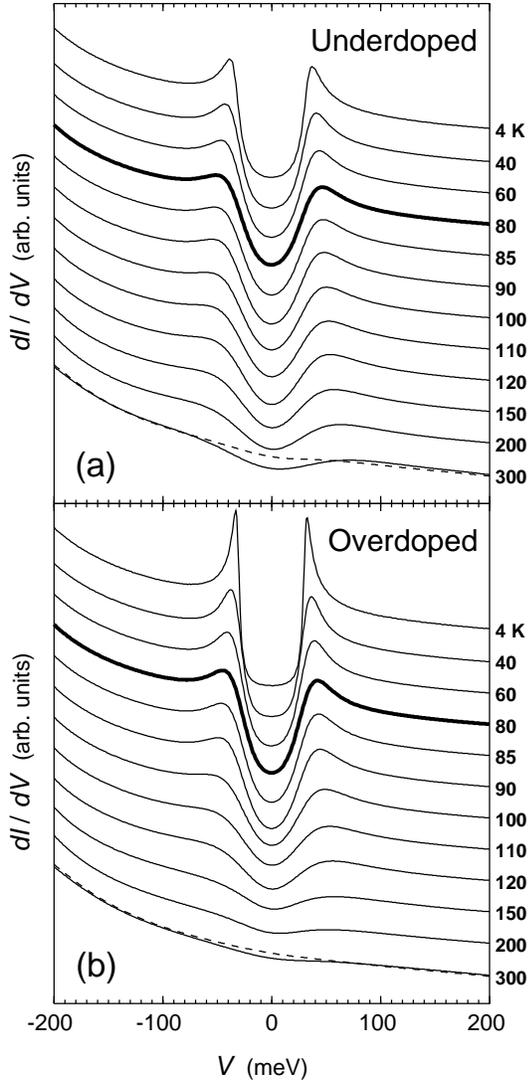}\end{center}
\caption{
Tunneling conductance as a function of temperature. The model parameters
represent (a) underdoped ($V_0B^0=15$~meV, $B^0/B^1=2$) and (b) overdoped
($V_0B^0=7$~meV, $B^0/B^1=0.5$) situations. The critical temperature is
$T_c=80$~K (bold line). The dashed lines show the $T=300$~K spectra
corresponding to $B^1=0$ and (a) $V_0B^0=7.5$~meV, (b) $V_0B^0=3.5$~meV. The
curves have been shifted for clarity.
}\label{fig:sts}
\end{figure}

Neglecting possible anisotropies of the tunneling matrix element as well as
$k_z$-dispersion effects, we calculate the tunneling conductance as the
convolution of the density of states with the derivative of the Fermi function.
The result is shown in Fig.~\ref{fig:sts} for various temperatures. In order to
focus on the effect of local superconductivity and phase fluctuations, we have
kept the model parameters independent of temperature: the whole temperature
dependence of the curves, in Fig.~\ref{fig:sts}, relates to the variation of
the correlation length $\xi$ and Fermi function with $T$. A better fit to the
experimental data could be obtained, in principle, by allowing the amplitudes
$B^0$ and $B^1$ to vary with temperature. This would not, however, change the
qualitative conclusions we wish to draw. In Fig.~\ref{fig:sts}(a), $B^0$ is
larger than $B^1$ while in Fig.~\ref{fig:sts}(b) $B^1$ is larger than $B^0$. In
the next section, we argue that these two typical cases correspond to
underdoped (UD) and overdoped (OD) situations, respectively. The spectra shown
in Fig.~\ref{fig:sts} reproduce some of the characteristic features observed
experimentally in BSCCO samples.\cite{Renner_98} Both UD and OD curves evolve
smoothly across $T_c$ into a pseudo-gapped spectrum, the peak-to-peak distance
remaining approximately temperature independent. Moreover, the coherence peaks
and the gap structure disappear more rapidly in the OD case as the temperature
is raised, which is also consistent with the experimental findings. The model,
however, is not able to account for a number of experimental observations, such
as the asymmetry in the temperature dependence of the positive and
negative-bias conductance peaks, or the dip structure recorded at $\sim
2\Delta$ below $T_c$. We also note that the model
Eqs.~(\ref{eq:L}) has $s$-wave symmetry. The calculated
spectra are therefore not expected to agree in details with experiment at low
energies.

According to our model, the local superconducting correlations responsible for
the high temperature pseudo-gap also have implications below $T_c$. In the
underdoped case, the local (incoherent) superconducting correlations broaden
the zero temperature density of states. The resulting conductance spectra have
small coherence peaks and a rounded line-shape around the fermi energy. In the
overdoped case, in contrast, the $T=0$ curve looks more like a ($s$-wave) BCS
spectrum.

As the temperature increases from zero to $T_c$, the density of states remains
unchanged in both UD and OD cases, and the temperature dependence of the
conductance spectra relates solely to the Fermi function. This behavior
persists above $T_c$ in the UD case owing to the dominant role of $B^0$ (which
is $T$-independent in our calculations). In the OD situation, on the contrary,
the gap fills in rapidly above $T_c$ as the contribution of $B^1$ disappears
due to increasing phase fluctuations; at elevated temperatures, only a weak
pseudo-gap due to $B^0$ remains. Fig.~\ref{fig:sts} also illustrates the effect
of the temperature dependence of the amplitudes $B^0$ and $B^1$. At room
temperature, $B^1$ is expected to vanish and $B^0$ is expected to be smaller
than at low temperature. Taking $B^1=0$ and
$B^0(300~{\text{K}})=B^0(0~{\text{K}})/2$, one obtains the dashed spectra in
Fig.~\ref{fig:sts}, which no longer exhibit a sizable pseudo-gap structure.

\begin{figure}[tb!]
\leavevmode\begin{center}\epsfxsize6cm\epsfbox{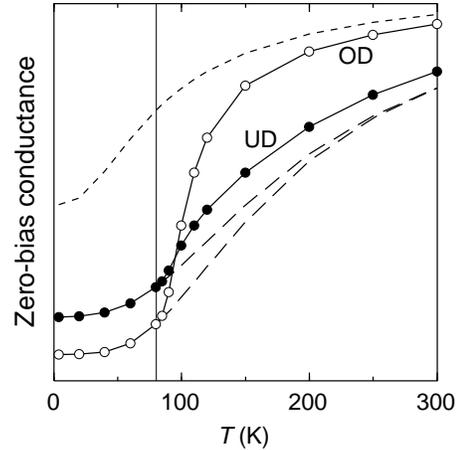}\end{center}
\caption{
Calculated zero-bias conductance as a function of temperature for underdoped
(UD, black symbols), and overdoped (OD, white symbols) systems. The parameters
are the same as in Fig.~\ref{fig:sts}. The dashed lines show the conductance
obtained by thermally broadening the $T=0$ spectra. The dotted line is obtained
by letting $B^1$ go to zero in the OD situation. The vertical axis starts at
zero.
}\label{fig:zero_bias}
\end{figure}

The difference between the temperature evolutions of the UD and OD spectra is
best seen in Fig.~\ref{fig:zero_bias}, where we plot the calculated zero-bias
conductance. Below $T_c$, the zero-bias conductance is larger in the UD case
due to strong local correlations. Above $T_c$, the conductance increases
sharply in the OD case, corresponding to the filling of the gap. In either UD
and OD cases, the zero-bias conductance above $T_c$ is larger than the value
expected by thermally broadening the $T=0$ spectra (see
Fig.~\ref{fig:zero_bias}).

From a general point of view, one can confirm from our calculations that the
sharpness of the peaks in the density of states (and correspondingly the size
of the zero-bias conductance) is related to the strength and range of the
superconducting correlations. The larger the ratio $B^1/B^0$ and/or the longer
the range $\varrho_0$, the sharper the peaks (the smaller the zero-bias
conductance). As an example, we show in Fig.~\ref{fig:zero_bias} the
conductance obtained by letting the coherence term $B^1$ go to zero in the OD
situation. Comparison of the curves with and without $B^1$ shows that the phase
coherence has the effect to depress the density of states at the Fermi energy
--- therefore raising the coherence peaks --- below $T_c$ and in some
temperature range above $T_c$, where the correlation length $\xi(T)$ is large.

\subsection{Angle-resolved photoemission}

\begin{figure}[tb!]
\leavevmode\begin{center}\epsfxsize8.5cm\epsfbox{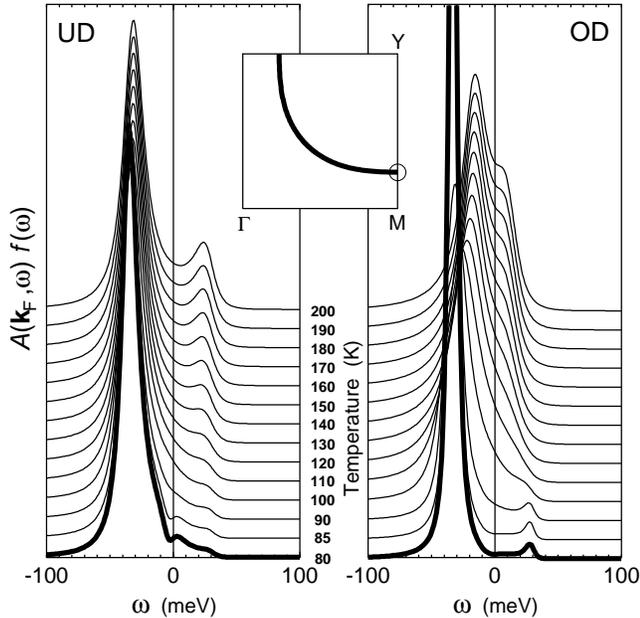}\end{center}
\caption{
Calculated ARPES intensity near $(\pi,\,0)$ as a function of temperature for
underdoped (UD) and overdoped (OD) systems. The parameters are the same as in
Fig.~\ref{fig:sts}. The curves have been shifted for clarity. Inset:
representation of the Brillouin zone showing the Fermi surface used in the
calculations and the Fermi crossing near $(\pi,\,0)$.
}\label{fig:arpes}
\end{figure}

Experimentally, it is found that the temperature dependence of the energy
dispersion curves measured by ARPES near $(\pi,\,0)$ depends on doping. In
overdoped samples, the leading-edge midpoint energy moves toward the Fermi
energy --- suggesting that the gap closes --- as $T$ increases above $T_c$. The
temperature variations of the midpoint energy are usually smaller in underdoped
samples.\cite{Hinks_98,Norman_98}

Apart from a matrix element, the ARPES intensity is just the product of the
spectral and Fermi functions. This quantity, calculated at the Fermi crossing
near the $M$ point, is shown in Fig.~\ref{fig:arpes} as a function of
temperature. A clear difference between the temperature evolution of the
spectral peak in the UD and OD cases can be seen. Consistently with experiment,
the peak shifts toward the Fermi energy in the OD case as $T$ increases. In the
UD case, the peak position is to first approximation independent of
temperature. Below $T_c$, the curves are almost identical to the spectrum at
$T_c$ (because the temperature $T<T_c\approx 7$~meV is small with respect to
the peak energy $\sim 35$~meV) and are not shown. One can see that the
quasiparticle peak is much sharper at $T_c$ in the overdoped as compared to the
underdoped system. This has also been seen experimentally\cite{Norman_98} and
can easily be understood in our model. The destruction of long range order by
phase fluctuations clearly affects qualitatively the spectral functions in the
OD case where the transition across $T_c$ is accompanied by a decrease of the
quasiparticle lifetime and increase of the intensity at the Fermi energy.

\begin{figure}[tb!]
\leavevmode\begin{center}\epsfxsize7cm\epsfbox{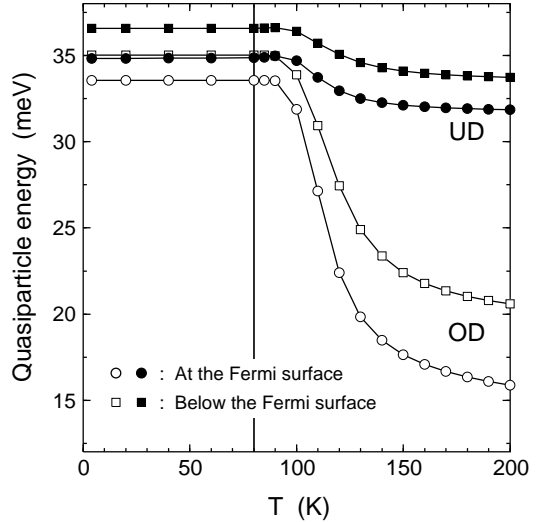}\end{center}
\caption{
Temperature dependence of the quasiparticle energy (maximum of the curves in
Fig.~\ref{fig:arpes}) at the Fermi crossing near $(\pi,\,0)$ for underdoped
(UD, black circles), and overdoped (OD, white circles) systems. The squares
show the energy of the quasiparticle at a $\bbox{k}$ point just below the
Fermi surface along the $M$--$Y$ line.
}\label{fig:gap_T}
\end{figure}

The position of the main quasiparticle peak in Fig.~\ref{fig:arpes} is reported
in Fig.~\ref{fig:gap_T} as a function of temperature. The temperature
dependence of the gap was studied in Refs~\onlinecite{Franz_98} and
\onlinecite{Norman_98} by fitting the experimental ARPES curves to a three
parameters Green's function. For overdoped samples, the gap was found to
decrease with increasing temperature (Ref.~\onlinecite{Norman_98}), in a way
very similar to what we obtain in the OD case, although the decrease was found
to begin already below $T_c$. Note that a small finite gap persists at all
temperatures in our calculations, since no temperature dependence of $B^0$ and
$B^1$ was taken into account. In a real situation, $B^0$ and $B^1$ would both
vanish at some temperature above $T_c$. In the underdoped samples, the gap was
found to be temperature independent within error bars
(Ref.~\onlinecite{Norman_98}) or slightly decreasing above $T_c$
(Ref.~\onlinecite{Franz_98}). The trend in Fig.~\ref{fig:gap_T} is similar. The
slight decrease of the gap above $T_c$ in the UD case results from the
suppression of the phase correlations as the temperature is raised.

The spectral line-shapes in Fig.~\ref{fig:arpes} are considerably sharper than
what is usually measured by ARPES, especially at elevated temperatures. The
estimated experimental resolution of $\sim 10$~meV cannot alone explain this
difference. Similar conclusions have been reached in
Ref.~\onlinecite{Franz_98}. It was shown there that the inverse quasiparticle
lifetime implied by fitting the experimental spectra are an order of magnitude
larger in ARPES with respect to STM. Inhomogeneities in the sample properties
could explain this discrepancy,\cite{Franz} since a much larger region of the
sample surface is probed by ARPES compared to STM.

The temperature dependence of the OD quasiparticle peak in Fig.~\ref{fig:gap_T}
contrasts with the apparent temperature independence of the gap width in
Fig.~\ref{fig:sts}. The coherence peaks in the density of states are due to
quasiparticle states with momenta just nearby $\bbox{k}_{\text{F}}$. Therefore,
one may expect that the energies of all these quasiparticles evolve in the same
way as the temperature increases. In this case, the coherence peaks would
rigidly follow this temperature evolution and both the STM and ARPES gaps would
close in the OD situation. We have found, however, that in our model the
energies of the quasiparticles at and nearby $\bbox{k}_{\text{F}}$ have
different temperature dependencies in the OD case. This is illustrated in
Fig.~\ref{fig:gap_T}, where we plot the energy of a quasiparticle with a
momentum $\bbox{k}$ just below the Fermi surface along the $M$--$Y$ line. For
$T<T_c$, this particular $\bbox{k}$ point contributes to the coherence peaks in
the density of states, since the corresponding energy is within $\sim2$~meV of
the energy at $\bbox{k}_{\text{F}}$. At 200~K, instead, the two energies differ
by $\sim5$~meV in the OD case, which is approximately half the width of the
zero temperature coherence peaks. This explains why the STM gap fills in
instead of closing although the ARPES gap at $\bbox{k}_{\text{F}}$ closes.
Thus, our results show that the apparent ``visual'' pseudo-gap may be different
in STM and ARPES data, even if each measurement is in agreement with the same
underlying theory.

\section{Conclusion}

Many workers in the field share the belief that the pseudo-gap phase in HTS is
a kind of mixed state, where strong short range superconducting correlations
coexist with long range phase disorder. This fact should reflect in the
properties of the Cooperon propagator, which should show ``partial''
superconductivity even above $T_c$. In this paper, we have shown that it is
possible to describe the properties of pseudo-gapped superconductors by writing
the superconductivity theory in general in terms of this Cooperon propagator,
and that reasonable phenomenological assumptions about the form of this
propagator lead to good agreement with experimental data. We have thus a
theoretical framework which is valid both above and below $T_c$, without
special treatment of the pseudo-gapped phase. Our main assumption is that the
relevant difference between overdoped and underdoped HTS is in the relative
magnitude of the short and long range parts of the Cooperon propagator,
described by the parameters $B^0$ and $B^1$, respectively. In the underdoped
HTS we assume that the ratio $B^0/B^1$ is larger than in the overdoped HTS. We
tentatively claim that $B^0$ is related to the single-particle energy gap
$\Delta_p$ measured by single-particle spectroscopy, while $B^1$ is related to
the coherence gap $\Delta_c$ measured in Andreev reflection or Josephson
experiments. As shown by Deutscher,\cite{Deutscher_99} $\Delta_p$ and
$\Delta_c$ differ in the HTS: the ratio $\Delta_p/\Delta_c$ is close to one in
the overdoped region and increases as the doping is reduced.

In this paper, we do not attempt to calculate the Cooperon propagator using one
or the other theoretical method. We first want to derive some empirical
constraints on the function $L$ from direct comparison with experiments. Our
approach is also limited, at this stage, to $s$-wave gap symmetry. We are
currently working on an extension of these calculations for $d$-wave symmetry
and on the calculation of the density of states in vortices. Also, comparisons
of our model with other detailed spectroscopic data below $T_c$ (vortices and
Josephson effect in particular) will show whether the approach presented here
is a fruitful one.

\begin{acknowledgements}

We wish to thank S.~E.~Barnes, H.~Beck, \O.~Fischer, M.~Franz,
B.~W.~Hoogenboom, and J.-M.~Triscone for very useful discussions.

\end{acknowledgements}


\begin{references}
%
%

\bibitem[*]{email}
Bernard.Giovannini@physics.unige.ch

\bibitem{Timusk_99}
For a recent review see, e.g., T. Timusk and B. Statt,
Rep. Prog. Phys. {\bf 62}, 61 (1999).

\bibitem{Renner_98}
Ch. Renner, B. Revaz, J.-Y. Genoud, K. Kadowaki, and \O. Fischer,
Phys. Rev. Lett. {\bf 80}, 149 (1998).

\bibitem{Roddick_95}
E. Roddick and D. Stroud, Phys. Rev. Lett. {\bf 74}, 1430 (1995).

\bibitem{Emery_97}
V. J. Emery and S. A. Kivelson, Nature {\bf 374}, 434 (1995).

\bibitem{Anderson_97}
P. W. Anderson, {\it The Theory of Superconductivity in the High-$T_c$
Cuprates\/}
(Princeton University Press, Princeton, 1997).

\bibitem{Randeria_89}
M. Randeria, J.-M. Duan, and L.-Y. Shieh,
Phys. Rev. Lett. {\bf 62}, 981 (1989).

\bibitem{Schrieffer_95}
J. R. Schrieffer and A. P. Kampf,
J. Phys. Chem. Solids {\bf 56}, 1673 (1995).

\bibitem{Chubukov_96}
A. V. Chubukov, D. Pines, and B. P. Stojkovi\'c,
J. Phys.: Condens. Matter {\bf 8}, 10017 (1996).

\bibitem{Lee_97}
P. A. Lee and X.-G. Wen,
Phys. Rev. Lett. {\bf 78}, 4111 (1997).

\bibitem{Ranninger_95}
J. Ranninger, J. M. Robin, and M. Eschrig,
Phys. Rev. Lett. {\bf 74}, 4027 (1995), and references therein.

\bibitem{Geshkenbein_97}
V. B. Geshkenbein, L. B. Ioffe, and A. I. Larkin,
Phys. Rev. B {\bf 55}, 3173 (1997).

\bibitem{Perali_99}
A. Perali, C. Castellani, C. Di Castro, M. Grilli, E. Piegari,
and A. A. Varlamov, cond-mat/9912363.

\bibitem{Loeser_96}
A. G. Loeser, Z.-X. Shen, D. S. Dessau, D. S. Marshall, C. H. Park, P.
Fournier, and A. Kapitulnik,
Science {\bf 273}, 325 (1996);
J. M. Harris, Z.-X. Shen, P. J. White, D. S. Marshall, M. C. Schabel, J. N.
Eckstein, and I. Bozovic,
Phys. Rev. B {\bf 54}, R15665 (1996).

\bibitem{Ding_96}
H. Ding, T. Yokoya, J. C. Campuzano, T. Takahashi, M. Randeria, M. R.
Norman, T. Mochiku, K. Kadowaki, and J. Giapintzakis,
Nature {\bf 382}, 51 (1996).

\bibitem{Curty_00}
For some recent efforts to separate the effect of phase fluctuations from size
fluctuations, see P. Curty and H. Beck,
Phys. Rev. Lett. {\bf 85}, 796 (2000).

\bibitem{Paramekanti_00}
A. Paramekanti and M. Randeria, cond-mat/0001109.

\bibitem{Kaminski_00}
A. Kaminski, J. Mesot, H. Fretwell, J. C. Campuzano, M. R. Norman, M. Randeria,
H. Ding, T. Sato, T. Takahashi, T. Mochiku, K. Kadowaki, and H. Hoechst,
Phys. Rev. Lett. {\bf 84}, 1788 (2000).

\bibitem{Renner_98b}
Ch. Renner, B. Revaz, K. Kadowaki, I. Maggio-Aprile, and \O. Fischer,
Phys. Rev. Lett. {\bf 80}, 3606 (1998).

\bibitem{Franz_98}
M. Franz and A. J. Millis, Phys. Rev. B {\bf 58}, 14572 (1998).

\bibitem{Kwon_99}
H.-J. Kwon and A. T. Dorsey,
Phys. Rev. B {\bf 59}, 6438 (1999).

\bibitem{Randeria_97}
M. Randeria, cond.mat/9710223.

\bibitem{Kadanoff_61}
L. P. Kadanoff and P. C. Martin,
Phys. Rev. {\bf 124}, 670 (1961).

\bibitem{Weiss_82}
L. Weiss and B. Giovannini,
Helvetica Physica Acta {\bf 55}, 468 (1982).

\bibitem{Janko_97}
B. Jank\'o, J. Maly, and K. Levin,
Phys. Rev. B {\bf 56}, R11407 (1997);
Q. Chen, I. Kosztin, B. Jank\'o, and K. Levin,
Phys. Rev. Lett. {\bf 81}, 4708 (1998);
Q. Chen, I. Kosztin, and K. Levin, cond-mat/9908362;
J. Maly, B. Jank\'o, and K. Levin,
Phys. Rev. B {\bf 59}, 1354 (1999);
I. Kosztin, Q. Chen, Y.-J. Kao, and K. Levin, cond-mat/9906180;
K. Levin, Q. Chen, and I. Kosztin, cond-mat/0003133.

\bibitem{Barnes_99}
S. E. Barnes, Int. J. of Mod. Phys. B {\bf 13}, 3478 (1999).

\bibitem{Guerrero_00}
M. Guerrero, G. Ortiz, and J. E. Gubernatis,
cond-mat/9912114.

\bibitem{Gupta_92}
R. Gupta and C. F. Baillie,
Phys. Rev. B {\bf 45}, 2883 (1992).

\bibitem{Schmalian_97}
J. Schmalian, S. Grabowski, and K. H. Bennemann,
Phys. Rev. B {\bf 56}, R509 (1997).

\bibitem{FFT}
The self-energy Eq.~(\ref{eq:sigma_k}) is most easily calculated using a Fast
Fourier Transform algorithm. To achieve a good precision, we set $0^+=1$~meV
and use a $1024\times1024$ $\bbox{k}$-point mesh in the Brillouin zone for
calculating the density of states, and a $2048\times2048$ mesh when calculating
individual spectral functions.

\bibitem{Hinks_98}
M. R. Norman, H. Ding, M. Randeria, J. C. Campuzano, T. Yokoya, T. Takeuchi, T.
Takahashi, T. Mochiku, K. Kadowaki, P. Guptasarma, and D. G. Hinks,
Nature {\bf 392}, 157 (1998); J. Phys. Chem. Solids {\bf 59}, 1888 (1998).

\bibitem{Norman_98}
M. R. Norman, M. Randeria, H. Ding, and J. C. Campuzano,
Phys. Rev. B {\bf 57}, R11093 (1998).

\bibitem{Franz}
M. Franz, private communication.

\bibitem{Deutscher_99}
G. Deutscher, Nature {\bf 397}, 410 (1999).

\end{references}
\end{document}